\documentclass[aps,prd,onecolumn,groupedaddress,showpacs,nofootinbib,amssymb
]{revtex4-2}
\usepackage[dvips]{graphicx}
\usepackage{amssymb}
\usepackage{amsmath}
\usepackage{graphicx,color}
\usepackage{amsfonts}
\usepackage{bm}

\begin{document}

\tolerance=5000
\title{Probing our Universe's Past Using Earth's Geological and Climatological History and Shadows of Galactic Black Holes}
\author{V.K. Oikonomou,$^{1,2}$}
\email{voikonomou@auth.gr;v.k.oikonomou1979@gmail.com;voikonomou@gapps.auth.gr}
\author{Pyotr Tsyba,$^{2}$}
\email{pyotrtsyba@gmail.com}
\author{Olga Razina,$^{2}$}
\email{olvikraz@mail.ru} \affiliation{$^{1)}$ Department of
Physics, Aristotle University of Thessaloniki, Thessaloniki 54124,
Greece}
\affiliation{$^{2)}$ L. N. Gumilyov Eurasian National
University, Nur-Sultan, Kazakhstan}

\begin{abstract}
In this short review, we discuss how Earth's climatological and
geological history and also how the shadows of galactic black
holes might reveal our Universe's past evolution. Specifically we
point out that a pressure singularity that occurred in our
Universe's past might have left its imprint on Earth's geological
and climatological history and on the shadows of cosmological
black holes. Our approach is based on the fact that the $H_0$
tension problem may be resolved if some sort of abrupt physics
change occurred in our Universe $70-150\,$Myrs ago, an abrupt
change that deeply affected the Cepheid parameters. We review how
such an abrupt physics change might have been caused in our
Universe by a smooth passage of it through a pressure finite-time
singularity. Such finite-time singularities might occur in
modified gravity and specifically in $F(R)$ gravity, so we show
how modified gravity might drive this type of evolution, without
resorting to peculiar cosmic fluids or scalar fields. The presence
of such a pressure singularity can distort the elliptic
trajectories of bound objects in the Universe, causing possible
geological and climatological changes on Earth, if its elliptic
trajectory around the Sun might have changed. Also, such a
pressure singularity affects directly the circular photon orbits
around supermassive galactic black holes existing at cosmological
redshift distances, thus the shadows of some cosmological black
holes at redshifts $z\leq 0.01$, might look different in shape,
compared with the SgrA* and M87* supermassive black holes. This
feature however can be checked experimentally in the very far
future.
\end{abstract}


\maketitle

\section*{Introduction}

The $H_0$-tension is an old problem in physics which still
troubles theoretical cosmologists, since no definite solution has
proven to solve it. Specifically, the tension on the $H_0$ value
exists between the large redshift sources like the Cosmic
Microwave Background (CMB) radiation \cite{Planck:2018vyg} and the
small redshift sources like the Cepheids \cite{Riess:2020fzl}.
Currently, the $H_0$ tension problem is quite timely and there
exists a large stream of research works which try to explain
theoretically the tension, see for example
\cite{Perivolaropoulos:2021jda,Perivolaropoulos:2022khd,Reeves:2022aoi,Verde:2019ivm,Vagnozzi:2021gjh,Perivolaropoulos:2021bds,Vagnozzi:2021tjv,Perivolaropoulos:2022vql,Odintsov:2022eqm,Odintsov:2022umu,Niedermann:2020dwg,Poulin:2018cxd,Karwal:2016vyq,Oikonomou:2020qah,Nojiri:2019fft,Mortsell:2021nzg,Dai:2020rfo,He:2020zns,Nakai:2020oit,DiValentino:2020naf,Agrawal:2019dlm,Yang:2018euj,Ye:2020btb,
Desmond:2019ygn,DiValentino:2019jae,OColgain:2018czj,Vagnozzi:2019ezj,
Krishnan:2020obg,DiValentino:2019ffd,Colgain:2019joh,Vagnozzi:2021gjh,Lee:2022cyh,Nojiri:2021dze,Krishnan:2021dyb,Ye:2021iwa,Yang:2018euj,Ye:2022afu,Verde:2019ivm,Marra:2021fvf}.
It is notable that the $H_0$-tension might be a problem of
calibration of the Cepheids
\cite{Mortsell:2021nzg,Perivolaropoulos:2021jda}, however no solid
proof on any aspect is given to date.

Recently it has be noted that an abrupt change in the global
physics of the Universe that occurred nearly $70-150\,$Myrs ago,
it might explain in an appealing way the $H_0$-tension problem
\cite{Perivolaropoulos:2021jda,Perivolaropoulos:2021bds,Perivolaropoulos:2022vql}.
Following this line of research, in
\cite{Odintsov:2022eqm,Odintsov:2022umu} we investigated how such
an abrupt global physics change might have been caused in our
Universe's past due to the occurrence of a pressure finite-time
cosmological singularity, also known as sudden or Type II
singularities. These cosmological singularities are rather smooth,
causing definable features in the Universe's recent past, which
could have had a direct impact on Earth's climatological and
geological history and also might have directly affected the
shadows of cosmological black holes corresponding exactly to this
redshift $70-150\,$Myrs ago. Pressure singularities are global
events in the Universe, because these singularities occur at a
finite time instance $t=t_s$ and occur globally on the
3-dimensional spacelike hypersurface at $t=t_s$. In this paper we
shall review these possible scenarios for our Universe's past and
we bring together all the material needed for this review.
Specifically, we shall show that these pressure singularities
naturally occur in modified gravity and specifically $F(R)$
gravity. Also, such finite-time singularities affect the orbits of
bound objects in the Universe. Thus, it might be possible that if
such a singularity occurred $70-150\, $Myrs ago, might have
affected Earth's elliptic orbit around the Sun, and in effect, the
climate on Earth might have changed globally, or even the strong
tidal effects might have changed abruptly the geology of Earth. On
the other hand, at larger scales, a pressure singularity might
have affected the photon orbits around cosmological supermassive
black holes, and in effect their shadows. In order to model this,
we shall use the McVittie metric
\cite{McVittie:1933zz,Faraoni:2007es,Kaloper:2010ec,Lake:2011ni,Nandra:2011ui,Nolan:2014maa,Maciel:2015dsh,Nolan:2017rtj,Perlick:2018iye,Perez:2021etn,Bisnovatyi-Kogan:2018vxl,Tsupko:2019mfo,Perez:2019cxw,Perlick:2021aok,Nojiri:2020blr}
because it is a consistent description for a black hole in an
expanding spacetime. Indeed, the Universe's expansion should
affect the cosmological black holes and the only consistent metric
that describes such an inhomogeneity in an expanding Universe is
the McVittie metric. Some time ago it was debatable whether the
McVittie metric truly described a black hole
\cite{Faraoni:2007es}, nowadays it is certain that the McVittie
metric describes a black hole in an expanding Universe
\cite{Kaloper:2010ec,Bisnovatyi-Kogan:2018vxl,Tsupko:2019mfo,Perlick:2021aok,Nojiri:2020blr}.
As we show, the photon orbits around supermassive McVittie black
holes are directly affected by a pressure singularity, thus this
may have definable features on their shadow. The shadows of
supermassive black holes are gradually developing to be quite
fruitful for new physics discoveries, see for example
\cite{Vagnozzi:2022tba,Vagnozzi:2022moj,Chen:2022nbb,Roy:2021uye,Khodadi:2020jij,Vagnozzi:2020quf,Allahyari:2019jqz,Bambi:2019tjh}
and references therein.

\section{Pressure Singularities in our Universe's Past and $F(R)$ Gravity}

Pressure singularities are rather smooth cosmological
singularities which are geodetically complete, and moreover the
strong energy conditions are respected during the passage of the
Universe through them. The classification of cosmological
finite-time singularities was made in \cite{Nojiri:2005sx}, so by
assuming that the cosmic singularity occurs at $t=t_s$,
finite-time singularities are classified as follows:
\begin{itemize}
\item Type I (``The Big Rip'') : This is a severe singularity of
crushing type at $t=t_s$, at which all the physical observable
quantities blow up, and specifically the pressure, the energy
density and the scale factor \cite{bigrip}. \item Type II (``The
Sudden Singularity''): This is the pressure singularity which we
shall consider in this review, see Refs.
\cite{barrowsudden,barrowsudden1} for details. In this case, the
scale factor and the energy density remain finite at $t=t_s$ while
the pressure diverges. \item Type III : This is also a crushing
type singularity, in which case the pressure and the energy
density diverge while the scale factor remains finite. \item Type
IV : This is a soft singularity in which case all the observable
quantities remain finite on the three dimensional spacelike
hypersurface defined by the condition $t=t_s$
\cite{Nojiri:2005sx,Nojiri:2004pf,Nojiri:2015fra,Odintsov:2015zza,Oikonomou:2015qha,Oikonomou:2015qfh},
and only the higher derivatives of the Hubble rate diverge, that
is $\frac{\mathrm{d}^nH}{\mathrm{d}t^n}\to \infty$, for $n\geq 2$.
\end{itemize}
In order to further understand what a pressure singularity is and
how this singularity can be realized by cosmological systems, let
us assume that the scale factor of the Universe has the following
form,
\begin{equation}\label{scalefactorini}
a(t)\simeq g(t)(t-t_s)^{\alpha}+f(t)\, ,
\end{equation}
with $g(t)$ and $f(t)$ are smooth functions of the cosmic time,
and the same applies for all their higher derivatives. For
consistency, let us assume that  $\alpha=\frac{2m}{2n+1}$, in
order to avoid having complex values for the scale factor, so
according to the classification of finite-time singularities
listed above, depending on the values of $\alpha$, we may have
these singularities for the scale factor (\ref{scalefactorini}):
\begin{itemize}
\item For $\alpha <0$ a Type I singularity is developed. \item For
$0<\alpha<1$ a Type III singularity is developed. \item For
$1<\alpha<2$ a Type II singularity is developed. \item For
$2<\alpha$ a Type IV singularity is developed.
\end{itemize}
The case we are interested in, that is the pressure singularity,
is developed for $1<\alpha<2$, since the first derivative of the
Hubble rate diverges $\dot{H}\sim \frac{(\alpha -1) \alpha
(t-t_s)^{\alpha -2}}{f(t)}$ and the same applies for the second
derivative of the scale factor $\ddot{a}\sim (\alpha -1) \alpha
g(t) (t-t_s)^{\alpha -2}$. As we already mentioned, for pressure
singularities, only the pressure diverges, and these singularities
are geodetically complete, since the following quantity is finite
\cite{Fernandez-Jambrina:2004yjt},
\begin{equation}\label{highercurvsc}
\int_0^{\tau}dt R^{i}_{0j0}(t)\, .
\end{equation}
More importantly, we have $\rho_\mathrm{eff}+3p_\mathrm{eff}>0$
and $\rho_\mathrm{eff}>0$, thus the strong energy conditions are
not violated when a pressure singularity occurs. In the context of
standard general relativity, the above conditions can be realized
by a scalar field and several cosmic fluids, but the systems must
be severely constrained. In the context of $F(R)$ gravity however,
the realization of the (\ref{scalefactorini}) and simultaneously
satisfying the strong energy conditions is much more easy due to
the fact that in the context of $F(R)$ gravity, a geometric fluid
does the job in an elegant and simple way. To have a direct grasp
on this, let us consider the $F(R)$ gravity action in the presence
of perfect matter fluids (cold dark matter and radiation),
\begin{equation}\label{action}
\mathcal{S}=\frac{1}{2\kappa^2}\int
\mathrm{d}^4x\sqrt{-g}F(R)+S_m(g_{\mu \nu},\Psi_m),
\end{equation}
with $\kappa^2=8\pi G$ and also $S_m$ stands for the action of the
perfect matter fluids. The field equations can be obtained by
varying the action with respect to the metric tensor in the metric
formalism, and these are,
\begin{align}\label{modifiedeinsteineqns}
R_{\mu \nu}-\frac{1}{2}Rg_{\mu
\nu}=\frac{\kappa^2}{F_R(R)}\Big{(}T_{\mu
\nu}+\frac{1}{\kappa^2}\Big{(}\frac{F(R)-RF_R(R)}{2}g_{\mu
\nu}+\nabla_{\mu}\nabla_{\nu}F_R(R)-g_{\mu \nu}\square
F_R(R)\Big{)}\Big{)}\, .
\end{align}
with $F_R(R)=\partial F(R)/\partial R$ and also $T_{\mu \nu}$
denotes the energy momentum tensor of the perfect fluids present.
The field equations can be cast in the Einstein-Hilbert form in
the following way,
\begin{equation}\label{geometridde}
R_{\mu \nu}-\frac{1}{2}Rg_{\mu \nu}=T_{\mu \nu}^{m}+T_{\mu
\nu}^{curv}\, ,
\end{equation}
with $T_{\mu \nu}^{m}$ being equal to,
\begin{equation}\label{tmnmat}
T_{\mu \nu}^{m}=\frac{1}{\kappa}\frac{T_{\mu \nu}}{F_R(R)}\, ,
\end{equation}
which originates from the perfect fluids, while the energy
momentum tensor $T_{\mu \nu}^{curv}$ originates purely from $F(R)$
gravity and it is equal to,
\begin{equation}\label{tmnmatcurv}
T_{\mu
\nu}^{curv}=\frac{1}{\kappa}\Big{(}\frac{F(R)-RF_R(R)}{2}g_{\mu
\nu}+F_R(R)^{;\mu \nu}(g_{\alpha \mu}g_{\beta \nu }-g_{\alpha
\beta}g_{\mu \nu })\Big{)}\, .
\end{equation}
What we effectively have at hand is an energy momentum tensor
acting as a source of gravity, which has a completely geometric
origin coming from $F(R)$ gravity terms.

For the consideration of the distortion of the planetary orbits,
we shall assume that the expanding Universe is described by a flat
Friedmann-Robertson-Walker (FRW) metric, with the following line
element,
\begin{equation}\label{frw}
ds^2 = - dt^2 + a(t)^2 \sum_{i=1,2,3} \left(dx^i\right)^2\, ,
\end{equation}
with $a(t)$ being the scale factor, and with the corresponding
Ricci scalar for the FRW metric being
\begin{equation}\label{ricciscalaranalytic}
R=6\left (\dot{H}+2H^2 \right )\, ,
\end{equation}
where as usual $H=\frac{\dot{a}}{a}$, stands for the Hubble rate.
In the same context, the field equations of $F(R)$ gravity for a
FRW metric, take the form,
\begin{align}\label{eqnsofmkotion}
& 3 H^2F_R=\frac{RF_R-F}{2}-3H\dot{F}_R+\kappa^2\left(
\rho_r+\rho_m\right)\, ,\\ \notag &
-2\dot{H}F=\ddot{F}_R-H\dot{F}_R +\frac{4\kappa^2}{3}\rho_r\, ,
\end{align}
with $p_r$, $\rho_m$ and $\rho_r$ being the pressure and the total
energy densities of the radiation and cold dark matter perfect
fluids. We further write the field equations in the
Einstein-Hilbert form as follows,
\begin{align}\label{flat}
& 3H^2=\kappa^2\rho_{tot}\, ,\\ \notag &
-2\dot{H}=\kappa^2(\rho_{tot}+P_{tot})\, ,
\end{align}
with $\rho_{tot}=\rho_{m}+\rho_{G}+\rho_r$ being the total energy
density, $\rho_m$ is cold dark matter energy density, and $\rho_r$
denotes the radiation perfect fluid energy density. Furthermore,
$\rho_{G}$ denotes the total energy density of the $F(R)$ gravity
geometrical fluid which is defined as,
\begin{equation}\label{degeometricfluid}
\kappa^2\rho_{G}=\frac{F_R R-F}{2}+3H^2(1-F_R)-3H\dot{F}_R\, ,
\end{equation}
and its pressure is,
\begin{equation}\label{pressuregeometry}
\kappa^2P_{G}=\ddot{F}_R-H\dot{F}_R+2\dot{H}(F_R-1)-\kappa^2\rho_{G}\,
.
\end{equation}
Effectively, the strong energy conditions are much more easily
satisfied in the context of $F(R)$ gravity, by the geometrically
generated perfect fluid with energy momentum tensor $T_{\mu
\nu}^{curv}$. It is useful to see explicitly what a pressure
singularity would imply for a flat FRW Universe. Firstly, before
and after the singularity, the Universe can superaccelerate or
superdecelerate for a slight amount of time near the singularity,
due to the fact that the second derivative of the scale factor
diverges at the time instance that the singularity occurs. This
feature complies with the abrupt changes in physics of Ref.
\cite{Perivolaropoulos:2016nhp}. Furthermore, another important
implication of the pressure singularity would be on the total
effective gravitational constant of $F(R)$ gravity. Considering
only subhorizon modes, that is modes which satisfy,
\begin{equation}\label{subhorapprx}
\frac{k^2}{a^2}\gg H^2\, ,
\end{equation}
the perturbations of matter density  defined by the parameter
$\delta =\frac{\delta \varepsilon_m}{\varepsilon_m}$, evolve
according to the following differential equation,
\begin{equation}\label{matterperturb}
\ddot{\delta}+2H\dot{\delta}-4\pi G_{eff}(a,k)\varepsilon_m\delta
=0\, ,
\end{equation}
with $G_{eff}(a,k)$ being the total effective gravitational
constant of  $F(R)$ gravity theory, which has the following form
\cite{Bamba:2012qi},
\begin{equation}\label{geff}
G_{eff}(a,k)=\frac{G}{F_R(R)}\Big{[}1+\frac{\frac{k^2}{a^2}\frac{F_{RR}(R)}{F_R(R)}}{1+3\frac{k^2}{a^2}\frac{F_{RR}(R)}{F_R(R)}}
\Big{]}\, ,
\end{equation}
with $G$ denoting as usual Newton's gravitational constant.
Obviously the terms containing the Ricci scalar and functions of
it, will directly be affected by a pressure singularity because in
a flat FRW spacetime, the Ricci scalar takes the form
$R=12H^2+6\dot{H}$, and $\dot{H}$ diverges at a pressure
singularity. Hence, this feature also complies with the argument
of Ref. \cite{Perivolaropoulos:2016nhp} of a sudden physics change
before 70-150Myrs which might have changed the Cepheid variables.
Also, the change directly on the effective gravitational constant
at small redshift as an explanation for the $H_0$-tension was also
mentioned in Ref. \cite{Marra:2021fvf}.

\section{Effects of a Pressure Singularity on Solar System Orbits and the Shadows of Galactic Black Holes}

Let us now discuss in a direct and quantitative way the
implications of a pressure singularity in the orbits of bound
objects and on the shadows of galactic black holes. We start off
with the former case, and in such a scenario, the pressure
singularity causes tidal forces in gravitationally bound objects
in the Universe, resulting in the disturbance of their orbits.
This scenario was considered sometime ago in Ref.
\cite{Perivolaropoulos:2016nhp}, and now we shall discuss the
implications of this scenario on the elliptic curve of Earth
around the Sun in our solar system and we shall also discuss the
effects of a disruption of Moon's orbit around Earth.
\begin{figure}
\centering
\includegraphics[width=18pc]{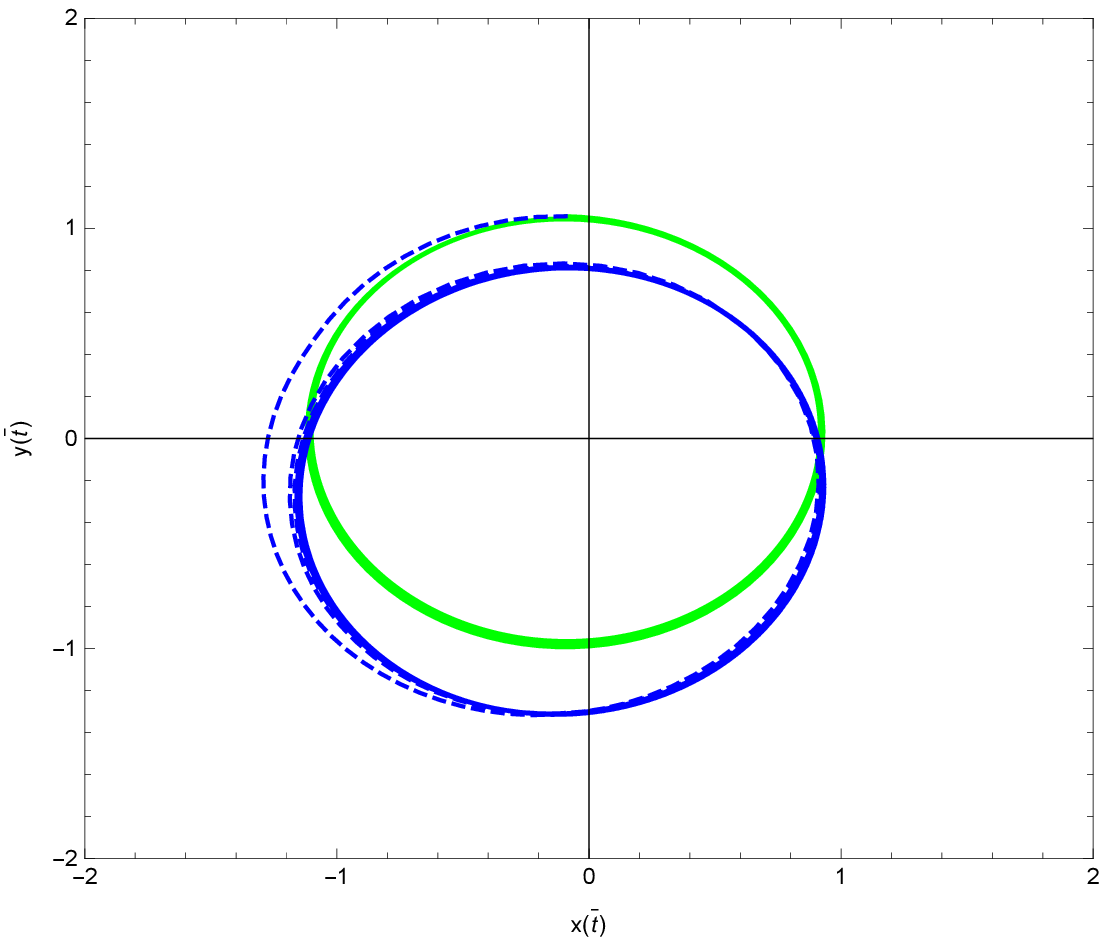}
\includegraphics[width=15pc]{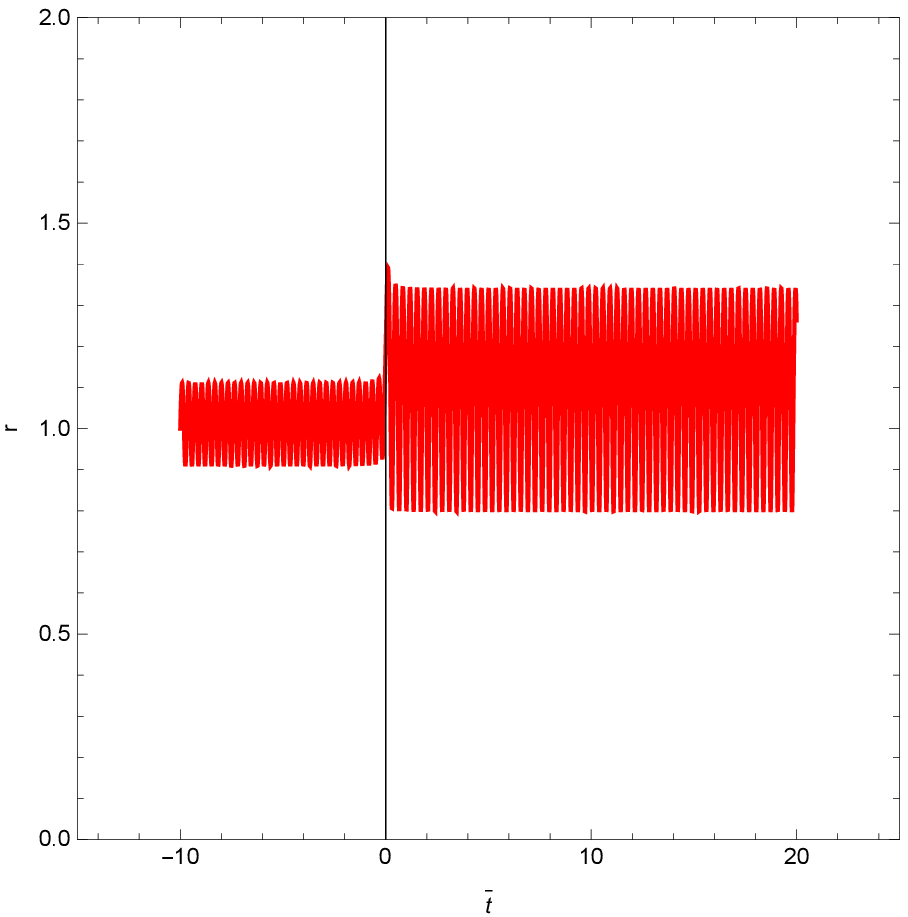}
\includegraphics[width=15pc]{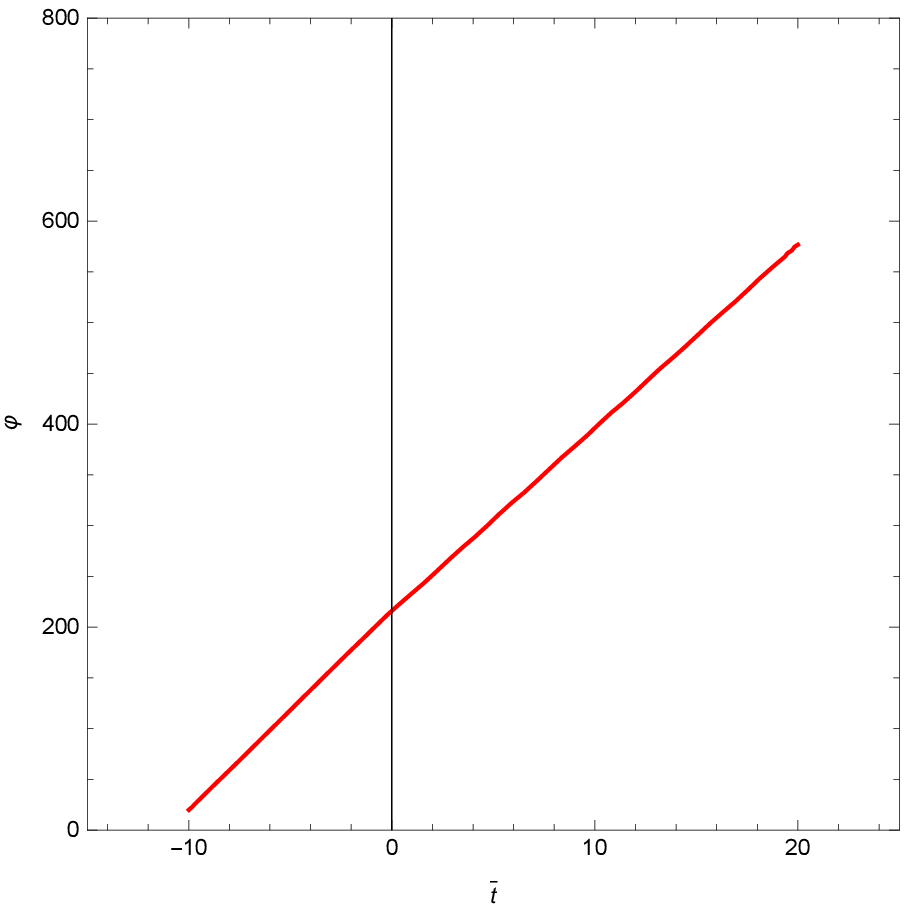}
\caption{ Upper left panel: Point particle trajectories in the
Cartesian plane before the pressure singularity (green curves)
during the pressure singularity and after the pressure singularity
(blue curves). Upper right panel: The functional dependence of the
radial coordinate $\bar{r}(t)$ with respect to the rescaled cosmic
time. Bottom Panel: the functional dependence of $\varphi(t)$ with
respect to the cosmic time.}\label{plot1}
\end{figure}
If this scenario indeed occurred $70-150\, $Myrs ago, then in
principle it would have dramatic effects on the climate and the
geology of Earth. Indeed, if Earth's orbit around the Sun was
disrupted before $70-150\, $Myrs, extreme temperature changes
would occur on the surface of Earth, even an ice age would be
justified, and we know that such an era actually occurred.
Accordingly, if the orbit of Moon was disrupted, the tidal effects
on the surface of Earth would be quite severe, and tides would be
so strong that perhaps entire seas would disappear from their
original places. In fact, the whole morphology of the surface of
the earth would abruptly change, and it is intriguing that the
morphological changes on Tethys Sea  occurred chronologically
nearly $70-150\, $Myrs ago.

Let us see in a quantitative way how the elliptic orbit could be
affected by the occurrence of a pressure finite-time singularity.
We shall use the framework developed in Ref.
\cite{Perivolaropoulos:2016nhp}, without invoking modified gravity
for simplicity. We start from a different orbit compared to Ref.
\cite{Perivolaropoulos:2016nhp}, and specifically from an elliptic
curve, instead of a circle. Following
\cite{Perivolaropoulos:2016nhp,Nesseris:2004uj}, for a FRW
Universe, the total energy and pressure are $\rho(t)=\frac{3}{8\pi
G} \left( \frac{\dot a^2}{a^2}\right)$ and $p(t)=\frac{1}{8\pi G}
\left( 2\frac{\ddot a}{a} +\frac{\dot a^2}{a^2}\right)$, hence it
is apparent that the pressure becomes singular for a pressure
singularity, due to the presence of the second time-derivative of
the scale factor. Let us model a pressure singularity in a simple
way, so assume that the singularity occurs at $t=0$ and therefore
the pressure singularity could occur if the scale factor has the
following form,
\begin{equation}
 a(t)=1+ c \vert t \vert^\eta\, . \label{scfactans}
\end{equation}
The values of the parameter $\eta$ for which the singularity
occurs are $1<\eta <2$. Consider a gravitational source with mass
$M$, the Newtonian limit of the metric describing the spacetime at
the vicinity of this compact object $M$ is
\cite{Perivolaropoulos:2016nhp,Nesseris:2004uj},
\begin{equation}
ds^2=\left(1-\frac{2GM}{a(t)\rho}\right)\cdot dt^2-a(t)^2\cdot
\left(d\rho^2+\rho^2\cdot (d\theta^2+sin^2\theta
d\varphi^2)\right)\, , \label{met}
\end{equation}
which is a correct description when $\frac{2GM}{a(t)\rho}\ll 1$
and note that this condition holds true even for pressure singular
 scale factors, since $a$ is not singular on pressure
 singularities. Introducing $ r=a(t) \cdot \rho $, the equations
 that determine the geodesics of the spacetime metric(\ref{met}) are \cite{Perivolaropoulos:2016nhp,Nesseris:2004uj},
\begin{equation}
(\ddot{r}-{\ddot{a}\over a}r)+{GM \over r^2}-r\dot{\varphi}^2=0,
\,\,\, r^2\dot{\varphi}=L\label{geodr}
 \end{equation}
where $L$ is the total angular momentum per unit mass, and it is
an integral of motion. Using equations (\ref{geodr}), we obtain
the radial equation of motion of a point particle around the
gravitational object with mass $M$
\cite{Perivolaropoulos:2016nhp,Nesseris:2004uj},
\begin{equation}
\ddot{r}={\ddot{a}\over a}r + {L^2 \over r^3}-{GM \over r^2}\, .
\label{radeqm1}
\end{equation}
It is vital to study dimensionless quantities, so we perform the
following rescalings ${\bar r}\equiv {r\over {r_0}}$, ${\bar
{\omega_0}}\equiv \omega_0 t_0$ and ${\bar t}\equiv {t\over t_0}$,
with $r_0$ and $t_0$ being arbitrary length and time scales. Upon
defining ${\dot \varphi}(t_0)= \omega_0\equiv {{GM}\over
{r_0^3}}$, the geodesics equation takes the following form
\cite{Perivolaropoulos:2016nhp,Nesseris:2004uj},
\begin{equation}
{\ddot {\bar r}}-{{\bar \omega_0}^2\over {{\bar r}^3}} + {{\bar
{\omega_0}}^2\over {{\bar r}^2}}-{{\ddot a}\over {a}}{\bar r}=0\,
. \label{dleqm}
\end{equation}
For a pressure singularity, the geodesics equation takes the
following form,
\begin{equation}
{\ddot {\bar{r}}}={{ \bar{\omega}_0}^2\over {{ \bar{r}}^3}} - {{
{\bar{\omega_0}}}^2\over {{
\bar{r}}^2}}+\frac{c\;\eta(\eta-1)\;\vert
\bar{t}\vert^{\eta-2}}{(c\;\vert \bar{t}\vert^{\eta}+1)}\bar{r} \,
.\label{sfsgeod}
\end{equation}
We performed a numerical analysis of the resulting geodesics
differential equation, giving various values to the parameters
$\eta$ and $c$, having in mind the disruption and not the
destruction of the bound system. Regarding the initial conditions,
we chose the initial orbit to be an elliptic curve. Our results
are presented in the three plots of Fig. \ref{plot1}. In the upper
left panel of Fig. \ref{plot1}, we present the trajectory of the
test mass around the massive object $M$, before the singularity,
during the pressure singularity and after the pressure
singularity. The green curve represents the elliptic orbit of the
test mass before the singularity, and at the vicinity of the
singularity, a disruption of the orbit occurs, and the orbit is
described by the blue curve, resulting to a final elliptic orbit
which is different from the initial orbit. In the upper right
panel we plot the radial coordinate of the orbit before and after
the singularity. Apparently, the two elliptic orbits are distinct
before and after the singularity. Finally, in the bottom panel of
Fig. \ref{plot1} we present the functional dependence of
$\varphi(t)$ with respect to the cosmic time. The same analysis
can be carried away for $F(R)$ gravity, but we refrain to analyze
this case, because the physics would be qualitatively identical to
the present case, and just the analysis would be slightly more
difficult to perform.

\begin{figure}
\centering
\includegraphics[width=18pc]{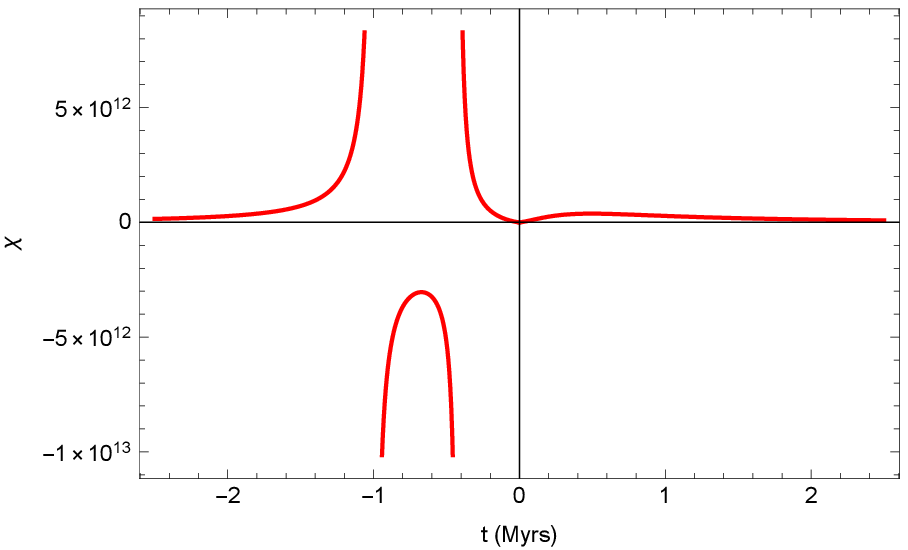}
\includegraphics[width=18pc]{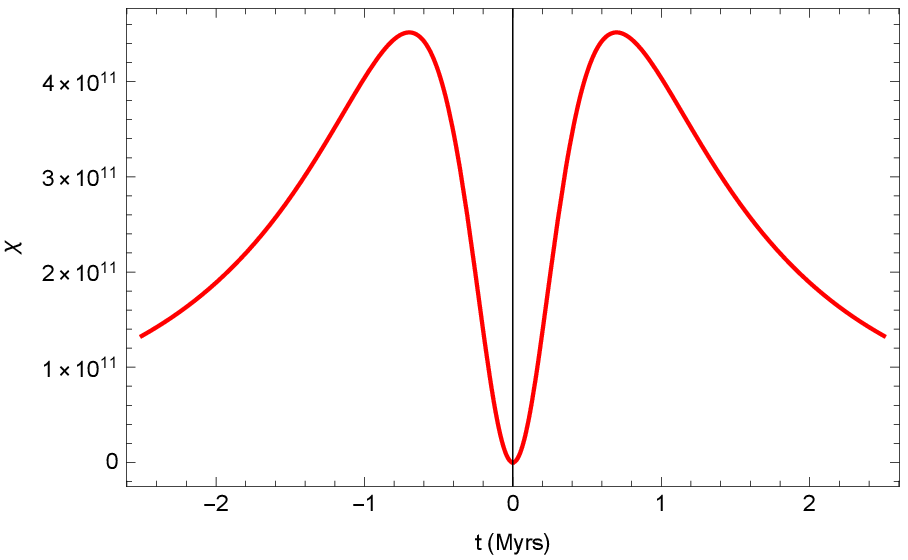}
\caption{The behavior of the quantity $\chi(R_c,t)$ as a function
of the cosmic time, by choosing $R_c=2.5\,m_0$, for the values
$\eta=5/2$ (left plot) and also for $\eta=2$ (right plot) and by
also choosing $c=1.5$ in Geometrized units. The plot in the left
panel depicts the case for which a sudden singularity occurs at
the time instance $t=0$. As it can be seen, the photon orbits do
not exist slightly before, during and after the sudden
singularity. The numerical analysis was performed for $2m_0<
R_c\leq 3m_0$ and the qualitative picture does not
change.}\label{plot2}
\end{figure}

Now let us discuss how the shadows of black holes may have
imprints of an abrupt physics change before 70-150 Myrs. For our
study we shall consider the McVittie metric
\cite{McVittie:1933zz,Faraoni:2007es,Kaloper:2010ec,Lake:2011ni,Nandra:2011ui,Nolan:2014maa,Maciel:2015dsh,Nolan:2017rtj,Perlick:2018iye,Perez:2021etn,Bisnovatyi-Kogan:2018vxl,Tsupko:2019mfo,Perez:2019cxw,Perlick:2021aok,Nojiri:2020blr}
which it is widely accepted today that it describes a black hole
in an expanding FRW Universe. The line element of the McVittie
metric in geometrized units ($G=c=1$) is,
\begin{equation}\label{mcvittiemetric}
ds^2=-\left(\frac{1-\frac{m(t)}{2r}}{1+\frac{m(t)}{2r}}\right)^2\cdot
dt^2-\left( 1+\frac{m(t)}{2r}\right)^4 a(t)^2\cdot
\left(dr^2+r^2\cdot (d\theta^2+sin^2\theta d\varphi^2)\right)\, ,
\end{equation}
with the function $m(t)$ being defined in the following way,
\begin{equation}\label{mfunction}
m(t)=\frac{m_0}{a(t)}\, ,
\end{equation}
with $m_0$ being the mass of the inhomogeneity or simply the
central body which is considered embedded in the expanding FRW
background. Basically, $m_0$ is the mass of the black hole, and
$a(t)$ is the scale factor of the Universe. Note that when $a=1$
the McVittie metric becomes identical to the Schwarzschild metric,
while when the mass of the central inhomogeneity reduces to zero,
the line element of the FRW spacetime is recovered. The best way
to investigate changes on the shadow of galactic black holes is
via studying the photon orbits around it. So now we will study the
photon orbits in the McVittie spacetime, and we shall focus on
geodesic paths on the plane determined by the condition
$\theta=\frac{\pi}{2}$. Due to the spherical symmetry of
spacetime, the angular momentum conservation condition yields,
\begin{equation}\label{conservationofangularmom}
\dot{\phi}=\frac{L}{R^2},\,\,\,\dot{\theta}=0\, ,
\end{equation}
with $R$ being the areal radius coordinate which is defined in the
following way,
\begin{equation}\label{arearadiuscoordinate}
R=a(t)r\left(1+\frac{m_0}{2 r a(t)} \right)^2\, .
\end{equation}
The geodesics equation of the circular photon orbits then reads
\cite{Perez:2021etn},
\begin{equation}\label{circulargeodesciscsphoton}
\frac{L^2}{R^2}=\left(f^2-g^2 \right)\dot{t}^2\, ,
\end{equation}
with the functions $f$ and $g$ being defined in the following way
\cite{Perez:2021etn},
\begin{equation}\label{functionsfandg}
f=\sqrt{1-\frac{2
m(t)}{R}},\,\,\,g=R\left(H+\frac{\dot{m}}{m}(f^{-1}-1)\right)\, ,
\end{equation}
where $H$ denotes the Hubble rate $H=\frac{\dot{a}}{a}$ as usual.
The physical quantity $\chi(R,t)=f^2-g^2=g^{\mu
\nu}\nabla_{\mu}r\nabla_{\nu}R$ stands for the definition of the
trapped spacetime regions and untrapped spacetime regions of the
total spherically symmetric spacetime. In order for the spacetime
to have circular photon orbits with radius $R_c$ for all the
cosmic time values, the following condition must hold true
\cite{Perez:2021etn},
\begin{equation}\label{stablecircularorbitscondition}
\chi(R_c,t)=f^2-g^2=1-\frac{2m(t)}{R_c}-R_c^2\left(H(t)+\frac{\dot{m}(t)}{m(t)}\left(\frac{1}{\sqrt{1-\frac{2m(t)}{R_c}}}-1
\right) \right)^2>0\, .
\end{equation}
When $\chi(t,R_c)<0$, the circular photon orbits do not exist, and
we will show, this is exactly what happens near a sudden
singularity. Let us take a simple form for the scale factor in
order to show this, so assume that,
\begin{equation}
 a(t)=c+ c \vert t \vert^\eta\, , \label{scfactans}
\end{equation}
with $c$ being an arbitrary constant with Geometrized units
$[L]^{-1}$ and also we shall make the assumption that
$\eta=\frac{2m}{2n+1}$ with the parameters $n$ and $m$ being
positive integers, in order to avoid having complex values for the
scale factor. The pressure singularity is developed at $t=0$ when
$1<\eta<2$, and we chose $t=0$ to be the singular point for
simplicity. Note that this singularity at $t=0$ might have
occurred at some time instance before $70-150\, $Myrs, and as we
will show, depending on the values of the free parameters $c$ and
$\eta$, the photon orbits with specific radii, and specifically
with radii $2m< R_c\leq 3m$, might not exist for a static black
hole in an expanding FRW background. We performed a numerical
analysis for the behavior of the quantity that determines the
existence or not of photon orbits, namely for $\chi(R_c,t)$ in
Fig. \ref{plot2}, by choosing the parameter $\eta$ to have the
values $\eta=2$ (right plot) and for $\eta=5/2$ (left plot), by
also choosing $c=1.5$ in Geometrized units.
As it can be seen in the left plot of Fig. \ref{plot2}, which
describes the case that the pressure singularity actually occurs
at $t=0$, the photon orbits exist well before the singularity at
$t=0$, but do not exist slightly before, during and slightly after
the singularity. In the right panel of Fig. \ref{plot2}, the
photon orbits exist before, during and after $t=0$, because there
is no pressure singularity in this case. We performed the same
analysis for several values of the photon orbit radii, and
specifically for $2m_0< R_c\leq 3m_0$, but the qualitative picture
does not change.

Now the question is what would the absence of photon orbits imply
for the shadow of galactic supermassive black holes. The latter
are at cosmological distances so they are situated at a non-zero
cosmological redshift. The absence of photon orbits would most
likely affect the ring of the galactic black hole shadow. Thus if
a large sample of galactic black holes is studied, corresponding
to redshifts $z\leq 0.01$, if there is a difference between the
shadows corresponding to small redshifts compared to the ones
corresponding to redshifts $z\sim 0.01$, this could probably
indicate that at high redshifts, some global physics change, like
a pressure singularity, might have occurred at the corresponding
redshift. However, observing a large sample of galactic black hole
shadows at larger redshifts, for the moment is a formidable task,
since there exist huge technical difficulties. Currently, the
observation of the M87 shadow and of the SgrA*
\cite{EventHorizonTelescope:2019dse,EventHorizonTelescope:2019ggy,EventHorizonTelescope:2022xnr}
is the best that we can get. There exists a large stream of
promising works for the shadows of black holes even corresponding
to these two galactic black holes, see for example
\cite{Vagnozzi:2022tba,Vagnozzi:2022moj,Chen:2022nbb,Roy:2021uye,Khodadi:2020jij,Vagnozzi:2020quf,Allahyari:2019jqz,Bambi:2019tjh}
and references therein. The M87 black hole exists at $z=0.004283$
thus due to the limitations on the resolution of VLBI techniques
at high redshifts, our proposal cannot be observationally verified
at present time, but rather it is a far future proposal.


\section{Conclusions}

In this short review we discussed a scenario for the late-time
Universe in which an abrupt physics change occurred $70-150\,
$Myrs ago, causing a change in the Cepheids variables and thus
resolving the $H_0$-tension problem. In our approach such an
abrupt physics change might have occurred due to a pressure
singularity which occurred $70-150\, $Myrs. The pressure
finite-time singularity is a smooth cosmological singularity of
timelike form, for which no geodesics incompleteness occurs, and
only the pressure diverges on the three dimensional spacelike
hypersurface defined by the time instance that the singularity
occurs. The strong energy conditions are not violated, and we
discussed the implications of such a singularity if this is
realized by an $F(R)$ gravity. Indeed, cosmological singularities
can easily be realized by $F(R)$ gravity and also the strong
energy conditions are also easily satisfied by the geometric fluid
generated by the $F(R)$ gravity. One of the notable effects of a
pressure singularity realized by $F(R)$ gravity, is the fact that
the effective gravitational constant blows up, and this feature is
perfectly aligned with the abrupt transition in physics argument
of Ref. \cite{Marra:2021fvf}. Notable and interesting is also the
fact that there possibly exists a connection between our approach
with the maximum turn around radius in the context of $F(R)$
gravity, see for example \cite{Capozziello:2018oiw}, see also
\cite{Capozziello:2014zda} for relevant cosmographic studies.

Assuming that a pressure singularity occurred in the recent past
of our Universe, we discussed in a qualitative way the effects of
this type of cosmological singularity on Earth's climate and
geological history and on the shadows of cosmological black holes.
Starting with the former scenario, if a pressure singularity
indeed occurred before $70-150\, $Myrs, it would globally affect
the orbits of compact objects. In such a scenario, both Earth's
elliptic curve around the Sun and the Moon's orbit around Earth
would be affected. In such a case, the distortion of Earth's orbit
around the Sun would case radical changes on Earth's climate
globally, causing in the most extreme scenario instant freezing on
the surface of the Earth. Accordingly, the distortion of Moon's
orbit around the Earth could cause tidal forces on the surface of
the Earth, which could reform the morphology of the surface of the
Earth, perhaps even dislocating entire seas.

Regarding the effects of a pressure singularity on the shadow of a
galactic black hole at a specific cosmological redshift, we
studied the existence of photon orbits around static black holes
in an expanding spacetime. As we showed, the photon orbits are
affected slightly before, during and slightly after the occurrence
of the pressure singularity. Thus if our scenario is indeed true,
this could have a measurable effect on cosmological black holes
which emitted light before $70-150\, $Myrs ago, thus at a specific
redshift. These supermassive galactic black holes would have
different shadows compared with the already mapped M87 and SgrA*.
However, our proposal cannot be verified at present time because
the resolution of the VLBI techniques forbid analyzing galactic
black holes at higher redshifts. In the far future however, such a
scenario could be analyzed in more detail for redshifts up to
$z\leq 0.01$ and there are further extensions of our simple
description we reviewed here. One should take into account the
fact that the shadow is a dynamical object, and is thus non-static
\cite{Bisnovatyi-Kogan:2018vxl,Tsupko:2019mfo,Perlick:2021aok,Nojiri:2020blr}.
The effects of the cosmic expansion on the diameter of the black
hole should also be taken into account \cite{Perlick:2018iye}.
Regardless that our proposal seem to belong to a far future to be
checked proposal, the current studies are interesting because
already up to the present time observations of high redshift
galactic black holes already exist \cite{Mortlock:2011va}, and it
is expected that these galactic black holes have large size
shadows \cite{Bisnovatyi-Kogan:2018vxl}. Perhaps also the James
Webb Space telescope can pin point properties of large redshift
supermassive black holes. The techniques used to obtain the
shadows of supermassive black holes are continuously developed
\cite{Younsi:2016azx,Abdujabbarov:2015xqa} and furthermore
theoretical and other aspects are also being refined
\cite{Vagnozzi:2022tba,Vagnozzi:2022moj,Chen:2022nbb,Roy:2021uye,Khodadi:2020jij,Vagnozzi:2020quf,Allahyari:2019jqz,Bambi:2019tjh,Addazi:2021pty,Miranda:2022brj}.

\section*{Acknowledgments}

This research is funded by the Committee of Science of the
Ministry of Education and Science of the Republic of Kazakhstan
(Grant No. AP14869238)

\end{document}